\documentclass[aps,prl,twocolumn,nofootinbib]{revtex4-1}
\usepackage{graphicx}
\usepackage{amsbsy} 

\newcommand{\VEV}[1]{\left\langle #1\right\rangle}
\begin{document}

\title{Proton size anomaly}

\author{Vernon Barger$^1$, Cheng-Wei Chiang$^{2,3,1}$, Wai-Yee Keung$^4$, and
  Danny Marfatia$^{5,1}$ \\}
\vspace{0.5cm}
\affiliation{%
\bigskip
$^1$ Department of Physics, University of Wisconsin-Madison, Madison, WI 53706,
USA \\
$^2$ Department of Physics and Center for Mathematics and Theoretical Physics,
National Central University, Chungli, Taiwan 32001, ROC \\
$^3$ Institute of Physics, Academia Sinica, Nankang, Taipei 11925, ROC \\
$^4$ Department of Physics, University of Illinois, Chicago, IL 60607, USA \\
$^5$ Department of Physics and Astronomy, University of Kansas, Lawrence, Kansas
66045, USA}

\begin{abstract}
A measurement of the Lamb shift in muonic hydrogen yields a charge radius of the proton that is smaller than the CODATA value by about 5 standard deviations. 
We explore the possibility that new scalar, pseudoscalar, vector, and tensor flavor-conserving nonuniversal interactions may be responsible for the discrepancy. We consider exotic particles that among leptons, couple preferentially to muons, and mediate an attractive
nucleon-muon interaction. We find that the many constraints from low energy data disfavor new spin-0, spin-1 and spin-2 particles as an explanation.
\end{abstract}

\maketitle


{\bf Lamb shift.} The success of quantum electrodynamics (QED) is apparent in the explanation of the Lamb shift~\cite{Lamb:1950zz} which is the observation that the 2S$_{1/2}$ state of hydrogen is higher than the 2P$_{1/2}$ state by about 1~GHz.\footnote{The dominant contributions to the Lamb shift arise from vacuum polarization and the vertex charge form factor of the lepton.  Vacuum polarization contributes negatively to $\Delta E \equiv E(2S_{1/2})-E(2P_{1/2})$ since more of the lepton's bare charge is revealed for the S state (than the P state) due to its greater overlap with the nucleus.  On the other hand, the vertex charge form factor is related to the zitterbewegung of the lepton which causes the effective Coulomb potential to be smeared out and less attractive. The effect is greater for the S state, so that the contribution to $\Delta E$ is positive.  The latter contribution is dominant for ordinary hydrogen but plays a minor role in muonic hydrogen because of the smaller
Compton wavelength of the muon. Consequently, $\Delta E$ is positive in ordinary hydrogen and negative in muonic hydrogen.}
Precision measurements in atomic spectra have tested bound-state QED to the extent that the charge distribution of the proton needs to be taken into account~\cite{Karshenboim:2005iy}.  The root-mean-square charge radius of the proton compiled by CODATA from  the spectroscopy of atomic hydrogen and electron-proton scattering is~\cite{Mohr:2008fa}
\begin{equation}
{\VEV{r_p^2}}^{1/2}=0.8768 \pm 0.0069~\rm{fm}\,,
\label{eh}
\end{equation}
provided there are no new long-range e--p interactions~\cite{Karshenboim:2010ck}.
It has been a long-held goal to measure the corresponding Lamb shift in muonic hydrogen which  is even more sensitive to the structure of the proton due to its smaller Bohr radius  $(\alpha m_\mu)^{-1}$, (where $\alpha \sim 1/137$ is the electromagnetic fine structure constant and $m_\mu \simeq 105$~MeV).  
Recently, the $2P_{3/2}^{F=2} \to 2S_{1/2}^{F=1}$ Lamb shift in muonic hydrogen was measured to be~\cite{Pohl:2010zz},
\begin{equation}
\Delta \tilde E \equiv E(2P_{3/2}^{F=2})- E(2S_{1/2}^{F=1})=206.2949 \pm 0.0032\ \rm{meV}\,,
\label{expt}
\end{equation}
while the predicted value is~\cite{Borie:1982ax,Pachucki:1996}
\begin{equation}
\Delta \tilde E=209.9779(49)-5.2262\VEV{r_p^2}+0.0347\VEV{r_p^2}^{3/2}\,,
\label{theory}
\end{equation}
where radii are in fm and energy in meV, and the number in parenthesis indicates the 1$\sigma$ uncertainty of the last two decimal places of the given number. (Note that $\Delta \tilde E$ is defined to be positive.)  Eqs.~(\ref{expt}) and (\ref{theory}) yield the order of magnitude more precise result~\cite{Pohl:2010zz},
\begin{equation}
{\VEV{r_p^2}}^{1/2}=0.84184 \pm 0.00067~\rm{fm}\,,
\label{muh}
\end{equation}
which is smaller than the CODATA value by about 5$\sigma$.  A partial resolution of the discrepancy may be found in a correlation between $\VEV{r_p^2}$ and the $r_p^3$-dependent third Zemach moment (since they contribute to the Lamb shift with opposite signs) and perhaps unreliable extractions of these from electron-proton scattering data~\cite{DeRujula:2010dp}. Nevertheless, a $4\sigma$ difference remains.  The possibility that the 4\% difference is a hint of a new gauge interaction with a natural scale $\alpha m_\mu$ has been entertained in
 Ref.~\cite{Marciano:2010}.  

In this Letter, we postulate the existence of a new interaction between muons and nucleons, and study its nature, bearing in mind the many experimental constraints.  The interaction must be attractive since $\Delta \tilde E$ measured in muonic hydrogen is larger than expected, signaling that the $2S_{1/2}$ state is subject to a stronger attraction than electromagnetic.

Scalar and spin-2 boson exchanges produce an attractive potential, giving positive contributions to $\Delta \tilde E$.  Pseudoscalar boson exchange is a derivative interaction involving the spins and velocities of the lepton and the nucleus, which becomes insignificant in the nonrelativistic limit, and irrelevant to the Lamb shift.  A boson with both scalar and pseudoscalar couplings violates CP conservation.  Such a scenario faces strong constraints from electric dipole moment measurements of leptons and nucleons, and is disfavored~\cite{Bernreuther:1990jx}.  Vector boson exchange (like photon exchange) can produce an attractive potential if the quantum numbers associated with the lepton and the nucleus are opposite in sign. Then Lamb shift phenomenology is like that of scalar exchange.  Axial-vector exchange couples the spins of the lepton and the nucleus in the nonrelativistic limit (with an effective potential $-\alpha_\chi ({\boldsymbol \sigma}_\mu \cdot {\boldsymbol \sigma}_p)e^{-m_\chi r}/r$~\cite{Karshenboim:2010cg}), and affects the hyperfine structure (but not the Lamb shift) so that the correction to the hyperfine splitting between the $2P_{3/2}^{F=2}$ and $2S_{1/2}^{F=1}$ levels for 
$m_\chi \gg \alpha m_\mu$ is 
$\alpha_\chi {\alpha m_r\over20} {\frac{1+10 {(m_\chi/(\alpha m_r))^2}}{(1+{m_\chi/(\alpha m_r)})^4}}$, in the notation defined below. However, since the axial-vector current is 
not conserved, the propagator gives a very singular contribution for $m_\chi \lesssim \alpha m_\mu$, 
which is unphysical.  Absent a well-defined model, we do not consider the axial-vector case any further.

\begin{figure}
\centering
\includegraphics[width=2.5in]{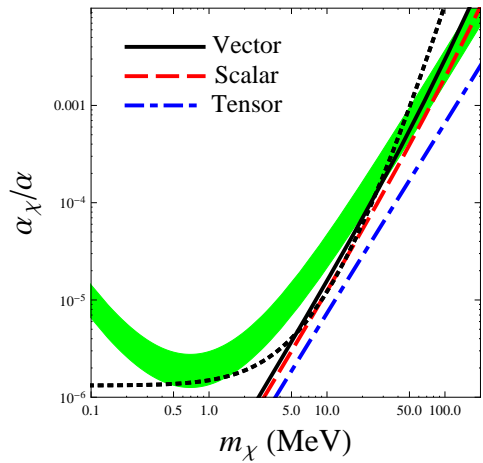}
\caption{\label{fig:lamb}%
The 95\% C.~L. range of $\alpha_\chi/\alpha$ required to reproduce the muonic Lamb shift is indicated by the green shaded region. The black solid, red dashed and blue dot-dashed lines 
are the {\it upper limits}  for vector, scalar and spin-2 particles, respectively, from a combination of 
n$-^{208}$Pb scattering data and the anomalous magnetic moment of the muon.
The black dotted curve is
 the upper bound obtained from atomic X-ray transitions.
 All bounds are at the 95\% C.~L.  
}
\end{figure}

Suppose the interaction between fermions $f$ and $\chi$ is given by $C_f^{S,V,T} \bar f f \chi$, where $S,V,T$ denote scalar, vector and tensor $\chi$, respectively, and $f$ can be a muon $\mu$ or a nucleon $n$; we assume isospin is conserved.  Throughout, we take the couplings $C$ to be real and positive. In the nonrelativistic limit, the muon-nucleon interaction is given by the Yukawa-type potential,
\begin{equation}
\label{eq:yukawa}
\Delta V(r)= -\alpha_\chi \frac{e^{-m_\chi r}}{r}\,,
\end{equation}
where $\alpha_\chi = C_\mu^{S,V,T} C_n^{S,V,T}/(4\pi)$ and $m_\chi$ is the mass of the particle $\chi$. Physical systems in which tensor interactions (by which we mean spin-2 exchange) are governed by a Yukawa potential allow the identification, $C_f^{T} \equiv C_f^{S}$. This will be valid for all our constraints except those that involve the anomalous magnetic moment of the muon $a_\mu \equiv (g_\mu-2)/2$.  The correction to the muonic Lamb shift is~\cite{Borie:1982ax,Jaeckel:2010xx}
\begin{equation}
\delta(\Delta \tilde E) = \alpha_\chi  m_\chi 
\frac{{m_\chi \over \alpha m_r}}
{2 \left( 1+{m_\chi \over \alpha m_r} \right)^4}\,,
\end{equation}
where $m_r$ is the reduced mass of the muon-proton system, and its use is
numerically important as $m_r$ is smaller than $m_\mu$ by more than 10\%.  
The green shaded region in Fig.~\ref{fig:lamb} shows the 95\% C.~L. region that accommodates
the difference between Eqs.~(\ref{eh}) and (\ref{muh}).
We do not consider $m_\chi > 10$~GeV since the required
 $\alpha_\chi$ becomes larger than 20$\alpha$, in the nonperturbative regime.
 
{\bf Upsilon decay.} 
 For scalar $\chi$ in the mass range $2m_\mu-9.3$~GeV, the nonobservance of  
 radiative decays of the $\Upsilon(2S)$ and $\Upsilon(3S)$ resonances: 
 $\Upsilon\to \gamma\chi, \ \chi \to \mu^+\mu^-$,
 strongly constrains the $\Upsilon-\chi$ coupling~\cite{babar}, which we expect to be no smaller 
 than $C_n^S$; for a Higgs-like $\chi$, the coupling is naturally 
 $\cal O$$(m_b/m_n)\times C_n^S$, while for a universal interaction, it should be
 $\cal O$$(C_n^S)$. We conservatively take the $\Upsilon-\chi$ coupling to be $C_n^S$. In obvious notation~\cite{Wilczek}, 
 \begin{equation}
{\frac{BF(\Upsilon \to \gamma \chi)}{BF(\Upsilon\to \mu^+\mu^-)}} = {\frac{(C_n^S)^2}{4\pi\alpha}}\bigg(1-{m^2_\chi\over m^2_{\Upsilon}}\bigg)\,.
\label{upsilon}
\end{equation}
Under our assumption that the branching fraction of $\chi \to \mu^+\mu^-$ is unity, the 
90\% C.~L. upper limit on $C_n^S$ ranges from 
$(0.94-9.4)\cdot 10^{-3}$~\cite{babar}, where the lower
end of the range corresponds to smaller $m_\chi$. The values of $\alpha_\chi/\alpha$ needed to explain the muonic Lamb shift constrain
$C_\mu^S$ to lie above $\cal O$(1), $\cal O$(10) and $\cal O$(100) for 
\mbox{$m_\chi \sim 2m_\mu$, 1~GeV} and 9~GeV, respectively, couplings which are too large.

A vector $\chi$ can mediate leptonic decays of spin-1 quarkonia.  Since the only lepton that $\chi$ couples to is the muon, one expects nonuniversality in leptonic decays.  For $\Upsilon$(1S) decays, $R_{\tau\mu} \equiv \Gamma_{\tau\tau} / \Gamma_{\mu\mu} = 1.005 \pm 0.013 \pm 0.022$~\cite{delAmoSanchez:2010bt}, whereas the SM expectation is $0.992$~\cite{VanRoyen:1967nq}.  The inclusion of $\chi$ modifies the SM value of $R_{\tau\mu}$ by a factor,
\begin{equation}
{\left[
\left( 1 \pm \frac{\alpha_\chi}{\alpha Q_b} \right)
- \left(m_\chi/m_{\Upsilon}\right)^2
\right]^2}
{\left[
1 - \left(m_\chi/m_{\Upsilon}\right)^2
\right]^{-2}}\,,
\end{equation}
where the $+$ ($-$) sign corresponds to destructive (constructive) interference and $Q_b$ is the electric charge of the $b$ quark.  For $m_\chi \alt 1$ GeV, a conservative 95\% C.~L. (one-sided)  upper bound on $\alpha_\chi / \alpha$ (assuming the SM and $\chi$ contributions destructively interfere) is 
$8.8 \times 10^{-3}$.  In the range \mbox{1~GeV $\alt m_\chi < m_{\Upsilon}$}, the upper bound becomes even more stringent, falling monotonically with $m_\chi$.  The mass of 
a vector $\chi$ is restricted to be less than about 230~MeV in order to explain the muonic Lamb shift. 

\begin{figure}
\centering
\includegraphics[width=2.5in]{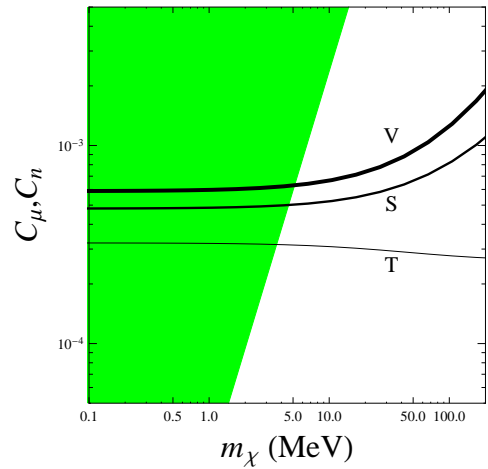}
\caption{\label{fig:anomaly}%
 The curves are 95\% C.~L. upper bounds on the muonic couplings 
 $C_\mu^S$, $C_\mu^V$ $C_\mu^T$ from $\Delta a_\mu$.  The green shading marks the values of the nucleon coupling $C_n^{S,V,T}$ {\it excluded} by n$-^{208}$Pb scattering at the 95\% C.~L. }
\end{figure}

{\bf Neutron scattering.}
Very precise neutron scattering experiments on heavy nuclei in the keV regime have
been performed to study the electric polarizability of the neutron. The goal is to measure
interference effects between the nuclear potential and the $r^{-4}$ potential produced
by electric polarizability. One can then see that a Yukawa potential 
$\mp A\,(C_n^{S,V,T})^2\,e^{-m_\chi r}/(4\pi r)$
may also be probed by such experiments; the minus and plus signs apply to  scalar/tensor and vector interactions, respectively. Stringent bounds are obtainable because the p-wave amplitude due to the short range strong interaction depends linearly on energy and differs markedly from
that due to the new longer range interaction. A n$-^{208}$Pb scattering experiment~\cite{Aleksandrov:1966} in the neutron energy range 1 to 26 keV measured the differential cross section (under the assumption that the scattering amplitude can be expanded in s and p waves) to be
\begin{equation}
{d\sigma}/{d\Omega} = {\sigma_0}
\left( 1 + \omega E \cos\theta \right)/(4\pi) \,,
\end{equation}
with $\sqrt{\sigma_0 / 4\pi} \simeq 10$~fm and $\omega = (1.91 \pm 0.42) \cdot 10^{-3}$ keV$^{-1}$. 
The measured values are in line with expectations so that the Yukawa potential contribution
ought to be subdominant.  Denoting the strong interaction contribution to $\omega$ by $\omega_s$, and
the contribution of the new interaction by $\Delta \omega$, clearly, $\omega =\omega_s + \Delta \omega$ 
with~\cite{Barbieri:1975xy}
\begin{equation}
\Delta\omega =
\mp \frac{16}{m_\chi^4} \frac{(C_n^{S,V,T})^2}{4\pi}\frac{A\,m_n^2}{\sqrt{\sigma_0 / 4\pi}}\,,
\end{equation}
in the Born approximation (not valid for $m_\chi \alt 0.1$~MeV), and $m_n$ is the neutron mass and $A$ is the atomic mass number. For scalar/tensor exchange, it is possible that a cancellation between $\omega_s$ and $\Delta\omega$ produces the experimental result. However, this cannot be the case for a vector $\chi$.  A conservative 95\% C.~L. (one-sided) upper limit can be obtained by 
requiring that $\Delta\omega \le 2.6\times 10^{-3} \mbox{ keV}$, {\it i.e.},
\begin{equation}
C^V_n \le ({m_\chi/206})^2\,,
\label{cvn}
\end{equation}
with $m_\chi$ in MeV. It is the shaded region of Fig.~\ref{fig:anomaly}.

While reliable bounds for a scalar/tensor $\chi$ are not extractable from the differential cross section, the total cross section measured between 10~eV and 10~keV~\cite{Schmiedmayer:1991zz, Leeb:1992qf} may be employed with confidence. The energy dependence of the $n-^{208}$Pb cross section for neutron energies below 10~keV can be parameterized by
\begin{equation}
\sigma(k) = \sigma(0) + \sigma_2 k^2 + {\cal O}(k^4) \,,
\end{equation}
where $k = 2.1968 \times 10^{-4} \sqrt{E} A/(A+1)$ is the wave vector of the incoming neutron (with $k$ in 
fm$^{-1}$ and $E$ in eV). The cross section in the limit of
vanishing momentum transfer $\sigma(0)$ is directly related to the scattering length, and $\sigma_2$
gives the effective range of the potential. The ${\cal O}(k)$ contribution to $\sigma(k)$
arises from the electric field of the nuclear charge distribution and is negligible. The measured values
$\sigma(0)=12.40\pm 0.02$~b and $\sigma_2=-448\pm 3$~b~fm$^2$ give a 95\% C.~L.  bound on $C_n^{S,T}$~\cite{Leeb:1992qf} that is almost identical to Eq.~(\ref{cvn}) in the mass range of
interest (and is not shown separately in Fig.~\ref{fig:anomaly}), but without the ambiguity from the cancellation mentioned above.

{\bf Muon anomalous magnetic moment.}
We now consider the independent constraint on $C_\mu$ from $a_\mu$.
  In fact, since the
experimental value of $a_\mu$ is above the SM expectation by more than three standard
deviations: $\Delta a_\mu \equiv a_\mu^{\rm exp} - a_\mu^{\rm th} = (29 \pm 9)
\times 10^{-10}$ \cite{Jegerlehner:2009ry},  the new interaction may
explain this difference. 
From Ref.~\cite{Leveille:1977rc},
\begin{equation}
\label{eq:S}
\Delta a_\mu =
{ {(C^{S,V}_\mu)}^2\over 8\pi^2}
\int_0^1 { 2x^2-\beta x^3 \over x^2+ (m_\chi^2/m_\mu^2) (1-x)}dx\,,
\end{equation}
where $\beta=1$ for a scalar and $\beta=2$ for a vector. For a tensor interaction, we trivially modify the result of Ref.~\cite{Graesser:1999yg}.  In the limit $m_\chi \ll m_\mu$,
\begin{eqnarray}
C^S_\mu &=& 4\pi({\Delta a_\mu/3})^{1/2} \ \ \alt \ \ 4.8 \times 10^{-4}\,,
\label{csmu}\\
C^V_\mu &=& 4\pi({\Delta a_\mu/2})^{1/2} \ \ \alt \ \ 5.9 \times 10^{-4}\,,
\label{cvmu}\\
C^T_\mu &=& 4\pi({3\Delta a_\mu/20})^{1/2} \ \ \alt \ \ 3.2 \times 10^{-4}\,,
\label{ctmu}
\end{eqnarray} 
where the one-sided upper bounds are at the 95\% C.~L.
From Fig.~\ref{fig:anomaly} it is evident that Eqs.~(\ref{csmu})-(\ref{ctmu}) apply for \mbox{$m_\chi \alt 10$~MeV}.

The bound in Eq.~(\ref{cvn}) can be combined with those in Eqs.~(\ref{csmu})-(\ref{ctmu}) to give the following 95\%~C.~L. constraints  for \mbox{$m_\chi  \alt 10$~MeV:}
\begin{eqnarray}
\alpha_\chi/\alpha &\alt& ({m_\chi/2847})^2\,    \ \ \ \ \        {\rm scalar}\,,\\
\alpha_\chi/\alpha &\alt& ({m_\chi/2573})^2\,     \ \ \ \ \       {\rm  vector}\,,\\   
\alpha_\chi/\alpha &\alt& ({m_\chi/3477})^2\,     \ \ \ \ \       {\rm  tensor}\,,                     
\end{eqnarray}
with $m_\chi$ in MeV. A similar (numerical) procedure can be applied 
for the entire range of $m_\chi$ to obtain
the upper bounds shown in Fig.~\ref{fig:lamb}. We see that a vector $\chi$ with mass between 25~MeV (with $ \alpha_\chi \simeq 10^{-4} \alpha$) and 210~MeV (with $\alpha_\chi \sim 10^{-2}\alpha$)
is a viable candidate. While a scalar $\chi$ with mass between 70~MeV and 210~MeV (with
$\alpha_\chi \sim (10^{-3}-10^{-2})\alpha$) is marginally allowed, a spin-2 $\chi$ is excluded.

{\bf Muonic atom transitions.} 
Measurements of the muonic
3D$_{5/2}-2$P$_{3/2}$ X-ray transition in $^{24}$Mg and $^{28}$Si atoms directly
constrain $\alpha_\chi$ for scalar, vector and tensor particles~\cite{Beltrami:1985dc}.  
For the Yukawa form of Eq.~(\ref{eq:yukawa}) with the coupling enhanced by a factor of $A$, the shift in the difference in energy levels from the QED expectation is~\cite{Beltrami:1985dc},
\begin{equation}
\frac{\Delta \cal{E}}{\cal {E}} =
\frac{2\alpha_\chi A}{5\alpha Z} \left[ 9f(2) - 4f(3) \right] \,,
\end{equation}
where $f(j) = [1 + j m_\chi / (2\alpha Z m_\mu)]^{-2j}$, $Z$ is the atomic number, and $j$ is the principle quantum number of the muonic state.  
The measured value obtained by averaging the results for $^{24}$Mg and $^{28}$Si, 
\mbox{$\Delta {\cal{E}}/{\cal{E}} = (0.2 \pm 3.1)\cdot 10^{-6}$}~\cite{Beltrami:1985dc}, gives the 95\% C.~L. bound (dotted curve) in Fig.~\ref{fig:lamb}. 
No additional area of the relevant parameter space is excluded by this constraint.

{\bf J/$\boldsymbol{\psi}$ decay.}
For $m_\chi < 2m_\mu$, the decay of scalar $\chi \to \mu^+\mu^-$ is kinematically forbidden so
that a constraint from the nonobservance of the  decay 
$J/\psi \to \gamma \chi$, with $\chi$ invisible~\cite{cleo}, may
be employed to exclude the marginally allowed region with  70~MeV~$<m_\chi <210$~MeV; preliminary data also exist for the  decay 
$\Upsilon(3S) \to \gamma \chi$~\cite{babar1}. A trivial modification of Eq.~(\ref{upsilon}) applies
to $J/\psi$ decay. The 90\% C.~L. upper limit on 
$BF(J/\psi \to \gamma \chi)$ is $\sim 4.5\times 10^{-6}$~\cite{cleo},
which when combined with $BF(J/\psi \to \mu^+\mu^-) =  (5.93 \pm 0.06)\cdot 10^{-2}$~\cite{pdg},
gives $C^S_n < 0.029$. Then, the muonic Lamb shift dictates that
$C^S_\mu$ be larger than $3.4\times 10^{-3}$ which is excluded at the 95\% C.~L.
by $a_\mu$; see Fig.~\ref{fig:anomaly}. Thus, scalars are also disfavored.

{\bf Pion decay.}  
The 90\% C.~L. experimental upper limit on the decay $\pi^0 \to \gamma \chi$, 
where $\chi$ is a vector particle, is $(3.3-1.9)\cdot 10^{-5}$ for $m_\chi$ ranging from
0 to 120~MeV~\cite{nomad}.
Equivalently,  $C^V_n < 4.5\times 10^{-4} (1- m_\chi^2/m_\pi^2)^{-3/2}$~\cite{Gninenko:1998pm},
and the corresponding values of $C^V_\mu$
 required to explain the muonic Lamb shift are excluded by 
$a_\mu$. This leaves $m_\chi$ between 120~MeV and  230~MeV.

{\bf Eta decay.}
 For vector $\chi$, the 90\% C.~L. experimental upper limit on invisible decays,
 $BF(\eta \to \chi\chi)/BF(\eta \to \gamma \gamma) < 1.65\times 10^{-3}$~\cite{bes}, translates
 into \mbox{$C^V_n \alt 0.05$}~\cite{Fayet}. The corresponding $C^V_\mu$ for 120~MeV$< m_\chi <  m_\eta/2 \sim 274$~MeV is excluded by $a_\mu$, so that vector $\chi$ is ruled out.

{\bf Conclusions.} 
We have found that new spin-0, spin-1 or spin-2 particles that mediate flavor-conserving nonuniversal spin-independent interactions are 
excluded by several low energy constraints as an explanation
of the proton radius anomaly.  We assumed that among leptons, the new particles couple only to the muon so as to avoid the large number of constraints involving the interaction of the electron with exotica. We also supposed that the coupling of the new particle to nucleons represents the minimal hadronic coupling, and employed it to mesons.

There are ways to relax some of the bounds at the expense of introducing complication. 
For example, since the contributions of scalars and pseudoscalars to $a_\mu$ 
are opposite in sign, allowing both a scalar boson and a pseudoscalar boson with
appropriately tuned couplings can lead to a cancellation that permits a rather large muonic
coupling. Then, although the hadronic couplings are highly restricted, the muonic Lamb shift
can be accommodated. Another possibility is that the new interaction violates 
isospin or $CP$, so that additional freedom is garnered.



%

{\it Acknowledgments.} 
This work was supported by DoE Grant Nos. DE-FG02-84ER40173, 
DE-FG02-95ER40896 and DE-FG02-04ER41308, by NSF
Grant No. PHY-0544278, by NSC Grant No. 97-2112-M-008-002-MY3, by NCTS, and by the WARF.

\end{document}